\documentclass[letterpaper,twocolumn,10pt]{article}
\usepackage{usenix}

\usepackage{cite}
\usepackage{amsmath,amssymb,amsfonts}
\usepackage{algorithmic}
\usepackage{graphicx}
\usepackage{textcomp}
\usepackage{xcolor}
\usepackage{macros}
\usepackage{comment}
\usepackage{url}
\usepackage{tikz}
\usepackage{booktabs}
\usepackage{enumitem}
\usepackage{tabularx,booktabs,wasysym,makecell,multirow,arydshln} 
\usepackage{rotating}
\usepackage{nicematrix}
\usepackage{subfig}
\usepackage{bigdelim}
\usepackage{adjustbox}
\usepackage{array}

\def\BibTeX{{\rm B\kern-.05em{\sc i\kern-.025em b}\kern-.08em
    T\kern-.1667em\lower.7ex\hbox{E}\kern-.125emX}}
\begin{document}

\title{``Did They F***ing Consent to That?’’:\\ Safer Digital Intimacy via Proactive Protection Against Image-Based Sexual Abuse}

\author{Lucy Qin \and Vaughn Hamilton \and Yigit Aydinalp \and Marin Scarlett \and Sharon Wang \and Elissa M. Redmiles}

\author{
{\rm Lucy Qin}\\
Georgetown University
\and
{\rm Vaughn Hamilton}\\
Max Planck Institute for Software Systems
\and
{\rm Sharon Wang}\\
University of Washington
\and
{\rm Yigit Aydinalp}\\
European Sex Workers Rights Alliance
\and
{\rm Marin Scarlett}\\
European Sex Workers Rights Alliance
\and
{\rm Elissa M. Redmiles}\\
Georgetown University
} 

\maketitle


\begin{abstract}
As many as 8 in 10 adults share intimate content such as nude or lewd images. Sharing such content has significant benefits for relationship intimacy and body image, and can offer employment. However, stigmatizing attitudes and a lack of technological mitigations put those sharing such content at risk of sexual violence. An estimated 1 in 3 people have been subjected to image-based sexual abuse (IBSA), a spectrum of violence that includes the nonconsensual distribution or threat of distribution of consensually-created intimate content (also called NDII). In this work, we conducted a rigorous empirical interview study of $52$ European creators of intimate content to examine the threats they face and how they defend against them, situated in the context of their different use cases for intimate content sharing and their choice of technologies for storing and sharing such content.
Synthesizing our results with the limited body of prior work on technological prevention of NDII, we offer concrete next steps for both platforms and security \& privacy researchers to work toward  safer intimate content sharing through proactive protection. 

\smallskip
\noindent\textit{\textbf{Content Warning:} This work discusses sexual violence, specifically, the harms of image-based sexual abuse (particularly in Sections ~\ref{sec:related} and ~\ref{sec:threats}).}
\end{abstract}


\section{Introduction}
Sharing intimate content\footnote{We define intimate content as images that: show a nude or semi-nude subject, contain intimate body parts, and/or intend to arouse. Throughout the text of this paper, we use the term \textit{images} to refer to photos and videos.} is a common practice. While prevalence is difficult to measure, estimates suggest that as many as 8 in every 10 adults share intimate content~\cite{traeen_gender_2022,falconer_body_2022,stasko2015reframing}. Sharing intimate content can benefit relationship intimacy~\cite{macdowall_sexting_2022,christine_geeng_usable_2020,waldman_law_2019,amundsen_hetero-sexting_2022,mcdaniel_sexting_2015,stasko2015reframing}, body positivity~\cite{waldman_law_2019,falconer_body_2022,hamilton2023nudes, bianchi_individual_2021}, and offer employment~\cite{hamilton2023nudes,sanders2023non,Redman}.

Societal norms and a lack of technological mitigations put people at risk of experiencing sexual violence when they share intimate content. Image-based sexual abuse (IBSA) encompasses a spectrum of violence, including (1) the nonconsensual distribution of consensually-created intimate content (NDII) \cite{said_NDII_2023} and the threat of NDII, (2) the nonconsensual creation of intimate imagery (e.g., via hidden cameras, ``upskirting,'' extortion, or AI)~\cite{mcglynn_IBSA_firstcoined_2017,henry_image-based_2022,eaton2023relationship,maddocks_deepfake_2020}, and (3) the unsolicited sharing of intimate content\cite{amundsen_male_2021, asia_eaton_2017_2017}. Studies estimate that one-third of individuals have been subjected to IBSA~\cite{powell_prevalence_2022} (similar to rates of other types of sexual violence \cite{CDC_SVprevalence_2022}). However, it is difficult to accurately estimate prevalence, as victim-survivors\footnote{We use the terminology victim-survivor to capture the range of ways in which people who have
experienced IBSA may identify~\cite{williamson_reconsidering_2018,sexual_assault_kit_initiative_victim_nodate,henry_image-based_2022}.} face significant stigma in reporting experiences and may not always be aware that content has been created or shared without their consent~\cite{henry_image-based_2022}.

We answer calls from prior work~\cite{christine_geeng_usable_2020,henry_image-based_2022,henry2018technology,henry2019responding,mcglynn2019shattering,waldman_law_2019} to \textbf{focus specifically on proactive technological mitigation of NDII}, leaving nonconsensual creation of images and post-harm mitigation as inquiries for future work. 
Through a qualitative interview study with $52$ Europeans adults who share intimate images for recreational (sexters) and commercial (sex workers) purposes we investigate the following as they relate to intimate content sharing:
\begin{itemize}[leftmargin=0.33cm]
 \setlength\itemsep{-0.33em}
    \item \textit{Contexts} (Section~\ref{sec:use:case}), which include group and 1:1 sharing with recipients that the sharer has an established relationship, a casual relationship, no relationship (strangers), and/or a commercial relationship.
     \item \textit{Threats}
    (Section~\ref{sec:threats}) such as recipient resharing (NDII) and non-recipient threats such as data breaches \& hacking, and unintentional discovery.
    \item \textit{Technologies} used to 
    share intimate content (Section~\ref{sec:sharing}) and how the contexts, threats, and other considerations we identify influence participant use of 40+ technologies including messaging platforms, social media, dating apps, fileshare platforms, email, and adult content platforms.
   
    \item \textit{Defensive strategies} against the threats we identify 
(Section~\ref{sec:strategies}), including detailed inquiry into the role of existing safety and security features.
\end{itemize}
We conclude in Section~\ref{sec:gaps-recommendations} by synthesizing the results of our study with the state of the art to offer a set of concrete recommendations to platforms and S\&P researchers on next steps to prevent NDII.

 \section{Related Work}
 \label{sec:related}

We summarize work related to the focus of our study: consensual \textit{and} nonconsensual distribution of intimate content.

\smallskip
\noindent\textbf{Technology \& Intimate Content.} 
Work exploring this topic from a technological viewpoint is scarce~\cite{henry2018technology}. Notably, 
Geeng et al. 
surveyed 247 U.S. adults quantifying their general sexting practices and security concerns~\cite{christine_geeng_usable_2020}. 
Respondents' sexting concerns included regret after sharing content, unwanted sexual content, NDII, and data breaches.
The authors found that their respondents managed their concerns by selecting platforms based on availability of specific features (e.g., expiring messages).
Additionally, they found evidence of behavioral strategies used to fill in the gaps left by platforms: respondents reported limiting what they share depending on the recipient and establishing rules with recipients. 

Relatedly, Waldman studied the behavioral mitigation strategies used by ``917 mostly gay male and bisexual users of geosocial dating apps'' and found that participants faced an implicit expectation to share intimate content in participating on the apps ~\cite{waldman_law_2019}. To mitigate risks, participants would delay sending images, avoid identifying features from photos, or select photos that they would be less embarrassed by in case of NDII.
Finally, examining trust in platform features that accommodate sexting use cases, Roesner et al. found that despite media focus on Snapchat's expiring messages feature for sexting, less than 2\% of their survey respondents used Snapchat primarily for sexting and fewer than 15\% used the app for sexting at all~\cite{roesner_sex_2014} because
they did not trust Snapchat nor its security practices and features.

We expand on these works, which provide a quantitative analysis of people's sexting practices and concerns, by using qualitative methods to focus on \textit{how} people use specific strategies to defend against NDII-related threats and \textit{why} and what gaps remain between their threat models and the defensive strategies available to them. 
In so doing, we answer Geeng et al.'s call for future work to ``identify—and bridge—any gaps between the threat models [that platforms] securely support and the threat models important to individuals engaging in sexting'' by systematically studying the ``technological properties'' offered by platforms; we additionally answer the call of criminology scholarship on NDII to specifically use qualitative methods to understand use of and potential for technological mitigation in this complex and nuanced space~\cite{henry2019responding}. 

Lastly, multiple works have explored user perceptions of security in messaging apps (often used for sharing intimate content)~\cite{abu-salma_obstacles_2017, akgul_evaluating_2021, dechand_encryption_2019, oesch_user_2022, justin_hendrix_what_2023, abu-salma_exploring_2018,wang_is_2023}. They found that users had large misconceptions from inconsistencies in (or lack of) information on end-to-end encryption (E2EE).
In lieu of reliable information, users may turn to folk theories to inform their decision-making \cite{wash_folk_2010} and develop a false sense of security or blanket mistrust.
Our findings are consistent with these results and we expand on this literature by including user perceptions of secure messaging in this context.
%
%
%
%

\smallskip
\noindent\textbf{Harms of IBSA.}
Existing literature on NDII is scarce. Thus, we include related work on both NDII and IBSA more broadly, with the understanding that NDII is a subcategory of IBSA. 

IBSA is a wrong in and of itself: ``regardless of the nature and extent of any further consequential harm...[IBSA] constitutes [a] breach of an individual’s \textquotesingle fundamental rights to
dignity and privacy, as well as their freedom of sexual expression and autonomy\textquotesingle ''~\cite{mcglynn2021s}. In the largest qualitative study of IBSA victim-surivivors to date ($n=75$),  McGlynn et al. describe IBSA as ``torture for the soul,'' specifically noting the devastating disruption to participants' lives, work, and selves as well as the permanence and ``endlessness'' of the abuse and its impacts
~\cite{mcglynn2021s}.

Beyond the wrong of IBSA alone, victim-survivors can experience harmful mental health consequences ranging from high levels of stress to suicidal thoughts and diagnoses of post-traumatic stress disorder, depression, and anxiety \cite{bates_revenge_2017,huber_harms_2023,asia_eaton_2017_2017}. IBSA victim-survivors are often subjected to further harms as they navigate the post-IBSA landscape by seeking justice, advice, legal assistance, and mental health support. A survey among 6,109 individuals from New Zealand, Australia, and the UK found that one-third of respondents expressed ``victim-blaming sentiments,'' defined as attributing blame towards victims and suggesting that they are at least partially responsible if an intimate image ends up online \cite{flynn_victimblaming_2023}. 
Victim-blaming has been associated with poorer outcomes for victim-survivors, acting as a barrier for reporting and help-seeking, worsening victim-survivors' already-impacted mental health \cite{mennicke_disclosure_2021}
\cite{orchowski_disclosure_2013}. Accordingly, approximately 73\% of IBSA victim-survivors choose not to report out of fear or embarrassment \cite{ruvalcaba2020nonconsensual}. 

Focusing on norms around perpetration, Hasinoff and Shepherd~\cite{hasinoff_sexting_2014} explored young adults' privacy norms related to sharing intimate images they receive, finding that while ``nearly all the respondents view maintaining privacy when sexting as an expected social norm,'' some felt it was acceptable to show intimate content they had received to others in person. Prior work finds that there is no one dominant motivation for IBSA perpetration; indeed perpetrators are often motivated to share by a desire to achieve social status -- with intimate images serving as a ``social currency'' -- rather than an intent toward revenge\footnote{We note that, while this is commonly referred to as ``revenge porn'' this term is not preferred by victim-survivors, activists, or researchers of IBSA~\cite{maddocks2018non}.} against a particular individual~\cite{henry_image-based_2022, asia_eaton_2017_2017}.

Literature on other digitally-mediated harms has identified NDII as a significant threat \cite{thomas_sok_2021}.
Prior work that addresses tech-facilitated intimate partner violence found that NDII was a common threat faced by victim-survivors~\cite{freed_stalkers_2018}. A study on digital abuse in South Asia found that women censor their presence on social media to avoid synthetic creation of intimate images using their likeness~\cite{sambasivan_they_2019}. Online harassment research finds that IBSA (and NDII) are growing concerns globally~\cite{schoenebeck_online_2023}. Prior work also illustrates how attackers exploit NDII to attack others~\cite{hutchings2019understanding}.
%

\smallskip
\noindent\textbf{Sex Work and IBSA.}
Sex workers are generally omitted from the discussion about IBSA, even though a large volume of intimate content is shared for commercial purposes. 
Waring describes sex workers as missing from  IBSA-related literature
due to academia, the law,\footnote{In England and Wales, for example, when content has been consensually uploaded online it ceases to be actionable as an offense if misused~\cite{sanders2023non}.} and even sex workers themselves regarding sex workers (like other victim-survivors) as at fault for uploading their own sexual imagery \cite{waring2021visual}.

In mixed-methods studies of commercial content creators who experienced IBSA, Sanders et al. and Redman and Waring found that 
participants were significantly affected 
when nonconsensually shared content
reached those they had hidden their sex work status from, outing them and putting them at financial and personal risk \cite{sanders2023non, Redman}. 
Clients or strangers weaponized this by threatening to share their content. Participants feared de-platforming when contacting platforms for assistance and reported a lack of response when they did so.
%
Hamilton et al. describe how the COVID-19 pandemic increased sex workers' digital exposure, including the volume of intimate content they share~\cite{hamilton2022risk}. Redman and Waring directly connect this with an increase in IBSA against sex workers. In separate work, Hamilton et al. document the growth in 
commercial content creation as a result of the pandemic and the mainstreaming of the OnlyFans platform~\cite{hamilton2023nudes}, underscoring the importance of studying both commercial and recreational sharers of intimate content, as we do in this work.

\smallskip
\noindent\textbf{Legal Perspectives.}
%
More laws are being introduced globally to address NDII \cite{cyber_civil_rights_initiative_nonconsensual_2023, uk_public_general_acts_section_2015, government_of_western_australia_criminal_2019, the_house_of_representatives_japan_act_2014,kirchengast_legal_2019}. While legislation is an important avenue for justice and changing societal perceptions~\cite{citron_fight_2022}
, legal scholars have described its many limitations. Legislation may evolve too slowly to keep up with ever changing digital threats or be too narrow to address international platforms
~\cite{citron_fight_2022}. Litigation is expensive and 
victim-survivors may not know who perpetrated the abuse, may fear retaliation, may not want to be in a courtroom with their abuser, or may not want to reveal their personal information to courts
~\cite{danielle_k_citron_sexual_2019}. As a result, it is critical to explore interventions on the technological side, especially those that platforms can implement~\cite{henry2019responding, mcglynn2019shattering}.

\section{Methodology}
\label{sec:methods}
We conducted 52 semi-structured interviews (August-November 2022) with adults (over 18 years old) living in European countries \cite{world_health_organization_countries_2023}
who had consensually shared intimate content in the past year: 24 with sex workers who produced their own digital content and 28 with those who shared for noncommercial purposes. 42\% of all participants were victim-survivors of IBSA (predominantly NDII).\footnote{One participant had experienced the nonconsensual \textit{creation} of intimate content.} 

\smallskip
\noindent
\textbf{Recruitment.} We recruited participants through convenience sampling via social media, relevant community organizations and personal contacts. To recruit recreational sexters, two authors posted a recruitment graphic to their social media accounts. To recruit sex workers, we used a similar but sex-work specific graphic (see Appendix~\ref{subsect:appendix-recruitment}). We contacted sex worker rights organizations to share it and the graphic was also posted in sex-worker only social media groups. 
The graphics included sign-up links that led to a Qualtrics survey that screened out individuals who are younger than 18 years old, had not sent intimate content in the past year, and/or were not  living in a European country. 
At the end of the screening survey, eligible participants were asked to review and sign a consent form for the interview study. \edit{We interviewed participants until we reached saturation for each group: recreational non-victim-survivors, recreational victim-survivors, commercial non-victim-survivors, commercial victim-survivors.}
To capture different perspectives,  participants were asked questions related to various dimensions of diversity in the screening survey (see Section~\ref{subsect:appendix-demos}).
Participants were compensated  €50 (or equivalent amount) through PayPal or Amazon giftcard.

\smallskip
\noindent
\textbf{Interview Data Collection.}
Four of the authors conducted the semi-structured interviews.
Sex-working participants were asked about their commercial \textit{and} recreational content, if applicable.
The interview protocols included several topics related to sharing intimate content (see Appendix \ref{appendix:questions}). We asked about the contexts in which participants shared intimate content (e.g., with whom, their motivations) and the technologies they used for storing and sharing intimate content. 
We then asked questions on how they navigate trust with recipients, security and safety concerns, and their proactive practices. Based on existing literature and media advice on sharing intimate content discussed in Section \ref{sec:related}, we selected a set of features that we explicitly asked about (listed in Appendix~\ref{appendix:questions}).
Lastly, we asked participants to imagine potential futures\cite{harrington_deconstructing_2019} for design changes and proposals for better technologies for intimate content. 
Although we asked victim-survivors  about their experiences with NDII, the \textbf{focus of this work is on proactive mitigation}.
We will analyze post-harm mitigation in follow-up work. 

\smallskip
\noindent
\textbf{Analysis.}
We followed an open coding analysis process. We randomly selected 4 interviews (2 commercial, 2 recreational) that three authors independently coded. They each created a codebook of main themes based on these interviews. Four of the authors (including the coders) then met to create the codebook.
As part of this process, we grouped smaller codes into larger themes. The coders then each randomly selected 4 interviews to code using the developed codebook. They reached ``substantial'' intercoder agreement \edit{(alpha= 0.72)} as measured by Krippendorf's Alpha~\cite{lombard2002content} and  
met to resolve any disagreements. 
They collectively 
finalized the codebook and then divided and independently coded the transcripts. Throughout the paper we provide a relative sense of the frequency of the themes we coded (e.g., that a theme was mentioned by half of participants) but emphasize the importance of not reducing ``sensitivity to complex concepts and
nuances in data'' by viewing these data through a quantitative lens~\cite{mcdonald2019reliability}.
%
%
%
%

\smallskip
\noindent
\textbf{Ethics.}
This work was approved by our institution's ethics review board. We took great care to protect the privacy of participants, which we detail in Appendix \ref{subsect:appendix-privacy}.
Since 42\% of the participants had experienced NDII, we warned all participants that discussing their experiences may be triggering and phrased questions to avoid re-traumatization.
To ensure both participant and researcher protection, our team includes a clinical psychologist who reviewed the interview protocols, analyzed data, and supported the research team. We additionally asked our partner organization, the European Sex Workers Rights Alliance (ESWA), and a victim-survivor of NDII for feedback on the protocol.
We followed trauma-informed best practices set forth by sexual violence researchers \cite{campbell_trauma-informed_2019} that minimize risk to participants and center victim-survivors’ choice, well-being, empowerment, and autonomy. Team members were trained in response techniques by a licensed clinical psychologist who specializes in sexual assault and trauma (e.g., offering to pause/stop the interview) and were prepared with resources (e.g., hotline information) to offer participants if needed.
We note that sharing these experiences may also be a form of empowerment for victim-survivors \cite{gundersen2021posting}.


\smallskip
\noindent
\textbf{Impacts and Research Justice.}
The interviews were transcribed by a sex worker (trained in transcription and compensated) as part of a research justice commitment to materially support communities we study. We published a white paper of this work through ESWA. We will share published research with participants who requested it and plan to develop a resource guide based on this work. \edit{In addition, we have been using these findings to advocate for platform changes and policies that may mitigate IBSA. We translated design recommendations from our participants into wireframes that we have presented to relevant platform product teams. We have also presented findings from this research to policymakers in the European Union, the United States, and Australia. Our partner organization, ESWA, has been using the white paper to conduct workshops on IBSA.}



\smallskip
\noindent
\textbf{Positionality.}
This work was conducted by a team of researchers who collectively are scholars of technology, sex work, and sexual violence.
This research focuses on adults living in Europe but only half of the research team is based in Europe. Given the diversity of our participants, our team does not completely reflect all of the genders, ethnicities, and sexual orientations of our participants. Therefore, there may have been some limitations in our interpretations of the data. 


\smallskip
\noindent
\textbf{Limitations.}
Since we conducted interviews in English and it is not the primary language in many European countries, this may have limited or skewed participation to those in countries where English is more prevalent and/or to those who are more educated, who tend to have studied English. 
Although we reassured participants that we were interested in understanding their perspectives to advocate for safer technologies for intimate content and took care to phrase questions non-judgmentally, the existing stigma of sharing intimate content may have prevented participants from fully disclosing their behaviors. 
Since there was a set of features we specifically asked participants about (Appendix~\ref{appendix:questions}), there may be more data on these features; however, each feature in this list was also independently mentioned by participants.


\smallskip
\noindent
\textbf{Participant Labeling.}
\edit{We intentionally did not distinguish between quotes from different participant groups and labeled all participants ``P\#.'' Given the stigmatization of sex work, we did not want quotes from sex workers to be dismissed or to be attributed as only relevant toward sex work (and we also asked sex workers about how they share intimate content in their personal lives). In general, there were many commonalities in the narratives shared across all participant groups and we offer Section~\ref{subsect:comparisons} for comparisons between them.} 

\smallskip
\noindent
\textbf{Platform Blinding.} With the exception of a few quotes, we replaced specific platforms with the type of platform mentioned in italics. A quote that says, ``I use WhatsApp'' would be replaced as, ``I use [\textit{messaging app}].'' This is to draw focus away from specific platforms and emphasize that \textit{intimate content is likely being shared across any and all platforms that enable the storing and/or sharing of images or videos.} 


\subsection{Demographics}
\label{subsect:appendix-demos}
\edit{When recruiting participants, we considered diversity across different dimensions. We present summaries of participant responses to various pre-interview survey questions below.}

\smallskip\noindent
\textbf{Geography.} Participants resided in 14 countries: Belgium, France, Germany, Greece, Ireland, Latvia, North Macedonia, Norway, Poland, Portugal, Serbia, Spain, The Netherlands, and the United Kingdom.

\smallskip\noindent
\textbf{Race.} 
Participants were given the option to self-describe their race through text-entry. The majority of  participants (65\%) identified as Caucasian/White (e.g., ``Caucasian,'' ``White, Russian'') including 2 mixed race individuals. Among those who did not (35\%), 15 opted to self-describe their race: 3 participants identified as Black (e.g., ``Black British,'' ``Black African''), 3 identified as Latina/Latino/Latinx, 2 identified as mixed race, and 2 identified as ``Brown'' or ``Southeast Asia, Brown.'' Other responses included ``French-Moroccan,'' ``...born in South America,'' ``Eurasian.''

\smallskip\noindent
\textbf{Education.} A majority (73\%) of participants had completed a Bachelor's degree or higher education. 17\% had completed some college, 6\% had completed high school (or equivalent), and 4\% did not finish high school.







\smallskip\noindent
\textbf{Disability.}
56\% of participants responded that they had a disability (physical, mental health, or neurodivergence) while 35\% did not (10\% did not answer). 

\smallskip\noindent
\textbf{Age.} Participants spanned the birth years of 2000 to 1976 with a median birth year of 1994 (roughly 28 years old at time of interview).

\smallskip\noindent
\textbf{Gender \& Sexuality.}
Participants were asked to optionally self-describe their gender and sexual orientation. 
To present summary statistics, we categorized the free-form text responses we received. These categories therefore represent a close approximation for some responses and do not capture the full richness of responses.

\begin{table}[!h]
    \centering
  \begin{tabular}{l|c}
    \toprule
    How would you describe your sexual orientation? & \% \\
    \midrule
    Asexual/Bisexual/Gay/Lesbian/Pansexual/Queer & 71\% \\
    Heterosexual & 21\%  \\
    Prefer Not to Answer & 10\% \\
    \bottomrule
    
    \toprule
    How would you describe your gender? & \% \\
      \midrule
Cis or Trans Woman & 27\% \\
    Cis or Trans Man & 38\%  \\
    Non-Binary, Gender Fluid, Agender & 19\% \\
    Prefer Not to Answer & 13\% \\
    Other & 2\%\\
    \bottomrule
  \end{tabular}
    \label{table:demographics}
\end{table} 
\section{Summary of Findings}
We conducted an interview study with 52 people in Europe who share intimate content (24 sex workers, 28 recreational sharers), 42\% of whom were victim-survivors of IBSA. Our participants used 40+ technologies to share intimate content. They relied on a combination of heuristics to choose technologies for sharing but nearly half chose to use their default method of communication.
Participants' threat models around NDII were focused on a recipient resharing their content but they were also concerned about data breaches, hacking, and unintentional content discovery. Echoing prior research,
all emphasized that stigma exacerbated these threats. 
To protect themselves, participants implemented a maze of safety strategies that included technical features (e.g., expiring messages) and interpersonal strategies (e.g., setting rules).
In sections \ref{sec:use:case} through \ref{sec:strategies}, we detail 
sharing contexts, threat models, how participants chose technologies for sharing content, and defensive strategies. In Section~\ref{sec:gaps-recommendations}, we include narratives on how technology could be improved to support digital intimacy/reduce NDII alongside recommendations for platforms and researchers.


\section{Sharing Contexts}
\label{sec:use:case}
Our participants shared intimate content across a wide range of use cases that we describe to contextualize their threat models and defensive strategies.
Participants shared \textbf{one-on-one} with a specific person and  with \textbf{groups} of individuals in all of the contexts detailed below.
The same individual may have multiple sharing contexts:
\begin{quote}
 ``I've shared them with a lot of people. I've shared them with romantic partners, I've sent them to strangers on the internet, like, on [\textit{social media}], I sometimes this search for people looking for [intimate content], and then just send them'' (P3).
\end{quote}


\smallskip
\noindent\textbf{Established relationship.}
A vast majority of participants shared intimate content in the context of an established relationship (e.g., spouse, long-term client).
In line with literature \cite{macdowall_sexting_2022, currin_motivations_2019}, participants frequently mentioned the importance of  intimate content as a form of connecting with their partners, ``intimate content is to keep a connection alive, it's part of a couple's sex life'' (P48).
Around a quarter of participants also shared content with friends as a form of bonding. ``I have been in text groups with like, female friends, and we've sent photos of our bodies to each other, just for fun, or like, hey, I'm feeling sexy today, or look at this great photo I took'' (P18).

\smallskip
\noindent\textbf{Casual relationship.} 
Participants also shared intimate content with short-term sexual partners. 
Many participants described sharing intimate content with someone they met on a dating app as means of initiating or maintaining a sexual relationship. 
``It's something that comes after a few basic questions that are asked in a regular [\textit{dating app}] conversation'' (P3).
Some shared in group contexts as well, such as ``very small groups where everyone was in sexual contact with each other'' (P2).

\smallskip
\noindent\textbf{Strangers.}
Individuals may also share intimate content with those they do not know and with whom they are not intending to initiate a relationship. For example, participants shared content mutually with strangers (e.g., in online communities) or in a one directional sharing relationship, such as with a social media audience. Our recreational participants were motivated by a desire to connect with others, express themselves artistically, receive body validation or express body positivity \cite{macdowall_sexting_2022,waldman_law_2019}. For example, P26 shared that, ``Early [in] my transition, I had the need to just share some pictures [to] feel more sexy...I created an account \textit{[on a social media platform]} especially for that, that [account] is networked with mostly other trans women that also have accounts for the exact same purpose.''

\smallskip
\noindent\textbf{Commercial.}
Sex workers in personal relationships shared in all of the contexts above. Additionally, their commercial sharing contexts closely mirrored the personal. They shared with: 
 \textit{established clients}; 
\textit{casual clients}---one-off clients or ``clients I do not trust (like new ones, or people I feel are asking me for a little too much without being vetted yet)'' (P33); and
\textit{unknown clients} such as those who bought their content from a subscription or clip site. 
Like the recreational users initiating hook-ups through dating apps, sex workers also shared content to verify themselves and entice further sales, whether of content or in-person services, like P37: ``because I'm doing escorting, I do share images with clients, not for sale, just to check me out.'' Finally, there were commercially-specific contexts like multi-service adult platforms: ``the one that I use, you can do live camming, and then you have your gallery of private pictures, and you set your price'' (P35).
\section{Threats}
\label{sec:threats}

Participants were concerned about NDII across different contexts. Primarily, they feared that a recipient would re-share their content.
Many also had additional concerns about non-recipient attackers via data breaches \& hacking as well as unintentional discovery (e.g., via device sharing, device theft/loss, or shoulder surfing).
Finally, participants noted a unifying mechanism of harm across these threats: stigma, which we discuss in further detail at the end of this section.

\subsection{Recipient Resharing}
\label{subsect:threats-recipient}
Despite many participants expressing that they trusted their recipients, they reported having passive or background fears about their content being reshared. P31 described,
``[in] that moment before [I] fall asleep, [I feel] like \textquotesingle Oh, shit,\textquotesingle~all of this is out there. [I] have a bit of a panic. But most of the time, I would say I feel relatively confident that if it does come out, I [will] deal with it in a way that I find suitable.''
Particularly for those who had experienced NDII previously, recipient resharing was a primary concern. A participant explained:
``I don't think there's ever been a time where I share content and I don't think about the harm that can be caused'' (P35).
Participants also reported secondary concerns such as not knowing they had been victimized: ``I know there are groups on Telegram and WhatsApp where people send porn. And then you don't have any control over it. 
Because if you're not part of these groups, you don't know they're there'' (P9). In addition, P3 noted the concern of aggregator pages further proliferating nonconsensual content by reposting it. News articles and research on the existence of anonymous forums and messaging groups that are dedicated to noncensually sharing intimate content support these fears 
\cite{semenzin_use_2020, henry_image-based_2019, bbc_disinformation_team_telegram_2022}.
Participants also spoke about fears of an ex-partner sharing intimate content after a breakup
\footnote{See \cite{coduto_delete_2024} for an in-depth analysis on how people manage intimate content after experiencing a breakup.}:
``I asked him to delete all my intimate images and videos, but I don’t know if he really did because we had a hard break up'' (P14).

%

Only five participants explicitly stated that they were not concerned about recipient resharing and NDII in general. A small number also expressed resignation toward the threat. Many sex-working participants expressed this privacy fatalism,  such as 
P42, who had been warned by peers: ``my colleagues in this told me, that you have to be prepared that in some point in your career, your photos will be in free pages...'' P39 added, ``it's something that I just accept.''  This sentiment was also shared by some recreational participants, particularly those who had shared material widely, with a large number of recipients or over a long period of time. P31 summarized this, saying, ``my ideal expectation is, of course they won't share it. But my realistic expectation is that it might happen anyway.'' In some cases, this led participants to adopt additional safety practices
such as P4 who frequently shared images with new recipients but, ``I've been sending the same three pictures since eight years'' or P24 who expressed that, ``it's very delusional to think that people have good intentions only so I only send what [I'm] comfortable sending and never show my face.''  We discuss security practices further in Section~\ref{sec:strategies}.



\subsection{Non-Recipient Threats}
\label{subsect:threats-non-recipient}
\smallskip
\noindent\textbf{Data Breaches \& Hacking.} Although many described hacking as ``a hypothetical background fear'' (P11), more than a third were concerned about it as a means of accessing 
their stored intimate content. They brought up risks on their personal devices (e.g. viruses, phishing
scams, and malware) and the risk of their 
intimate content being leaked through a data breach or poor security practices/policies of a platform they had used for sharing (e.g. how content is stored, protected, deleted). People feared both technical and policy breaches; exemplifying the latter, P10 ``read stories of [employees] who have set up programs to find [images] of women shirtless.''
\subsubsection{Unintentional Discovery}
\label{subsect:threats-unintentional}



\smallskip
\noindent 
\textbf{Device/Account Sharing.}
Almost a third of participants were concerned about others accidentally discovering intimate content while device sharing or using shared 
accounts.
For example, P22 commented, ``I have cousins, my mom, my grandma [around me at home]...they can take my phone anytime and browse [my photo] gallery.''
Similarly, P4 stores content on a file share platform and "sometimes I give [my parents] access to some of my [\textit{file share}] folders... So I make sure all the folders are well separated from each other. But I am always thinking twice when I'm giving access."

\smallskip
\noindent\textbf{Shoulder Surfing.}
About a sixth of participants worried about shoulder surfing: others accidentally viewing intimate content that is visible on a sender's or recipient's screen.
As P11 put it, ``have I been concerned that someone has accidentally seen a photo because I've sent it at the wrong time? 100\%.''
P12 described taking proactive action to prevent shoulder-surfing: ``I bump [an image] up [by sending text messages after the image], so that it wouldn't necessarily be the first thing that comes up on a screen.''

\smallskip
\noindent\textbf{Device Theft or Loss.}
More than a sixth feared that someone could access stored intimate content if a device were stolen or lost.
This was often coupled with concerns of weak password security: 
``someone like could crack my code and take my pictures'' (P30).
P51 experienced such an incident and described, ``once my phone was stolen, containing a lot, a lot, a lot of personal images and data...
I was like, \textquotesingle Oh, my [gosh]. Oh, my [gosh]\textquotesingle.''
Additionally, P38 asked a recipient to delete any intimate content prior to bringing a phone in for repair---a concern confirmed by cases of intimate content theft by customer service and technical support employees who were given device access \cite{gabrielle_fonrouge_rise_2022, gabrielle_fonrouge_t-mobile_2023, cnn_newsource_bank_2021}.







\subsection{Mechanism of Harm: Stigma}
IBSA can lead to many forms of harm (see Section~\ref{sec:related}) that stigma and victim blaming 
can perpetuate. 
Due to stigma, participants feared awkward or even dangerous situations with family, friends, their communities, and/or places of work and study. P18 worried about the reactions of their colleagues in the event of NDII:
``I would like to think that it would be a non issue, and the focus will be on the person who was nonconsensually sharing images of me. But unfortunately, there is a lot of victim blaming.''

Consistent with prior work~\cite{waring2021visual, sanders2023non,henry_image-based_2022, zara_ward_revenge_2022, asia_eaton_2017_2017}, the concerns described tended to depend on participant identity. For LGBTQ and sex-working participants, the risk was magnified if they or their partner were not out, or their identity was a personal risk to them or their recipient(s). 
For example, P10 took more precautions because,
``[My ex-partner] was from a quite conservative religious family... 
it's always better to play safe, especially when your partner is closeted."
In the context of sex work, P49 explained: ``I'm afraid that if I do more content... that someone from my family would see that... It's the most stressful thing for me,''
and as P35 described, ``I have another [non sex work] job
and I [get] really scared that I will no longer get work... 
[despite my] paranoia, there's nothing I can really do about it, because I have to still do [sex work].'' 
Cultural norms also magnified risks:
``Turkey is a very conservative and traditional country. My parents are open-minded but I don’t think they could handle this'' (P14).

In contrast, several cisgender men felt that they faced fewer threats\footnote{We note that men are disproportionately affected by sextortion, another form of IBSA, in which individuals are targeted by fake online personas to begin a relationship. Once intimate content is exchanged, the content is used as leverage for financial extortion~\cite{zara_ward_revenge_2022}.}: ``as someone who presents as male, I think I'm less at risk from exposure than someone who presents as anywhere else on the gender spectrum,
because I think my image is less likely to be shared around maliciously'' (P2).
Some also felt that if sharing intimate content were less stigmatized in their specific social community, they may receive more support if NDII occurred
(though they may still have concerns about the response of those outside of that community).
P26 commented that due to shared politics, they trusted ``people at my workplace to not let [nonconsenusally shared intimate content] have a negative influence."




\begin{table}[!b]
\footnotesize
\begin{tabular}{p{2cm}|p{4.25cm}|p{1cm}}
\toprule
\textbf{Sharing Method} & \textbf{Sharing Use Case}  & \textbf{\# of Platforms}\\
\toprule
Messaging &  1-1 conversation or with groups of recipients & 9 \\


\midrule
Social Media & 1-1 conversations (DMs), groups (posting or DMs), for marketing (commercial)  
&  10 \\
\midrule
Adult content platforms & Posting content to platforms, livecamming &  11 \\
\midrule
Dating / hookup apps & Sharing in-app intimate content, posting semi-nude/lewd photos in profile  & 10 \\
\midrule
Cloud/file share platforms & Storing and directly sharing content with recipients & 4 \\
\midrule
Email & Sharing larger files and/or links to files, occasionally used as an alternative to primary sharing technology & N/A \\



\bottomrule
\end{tabular}
\caption{Categories of sharing methods, use cases, and the number of platforms mentioned per category.}
\label{tab:sharingmethods}
\end{table}


\section{Technologies}
\label{sec:sharing}


\edit{In total, participants mentioned using 40+ different platforms for sharing and storing intimate content. We categorized them in Table~\ref{tab:sharingmethods}, which also includes the use cases people mentioned in relation to each category of sharing method and the number of different platforms that were mentioned per category. This leads us to emphasize that any platform that enables the storing and sharing of photo/video content is likely being used for intimate content.  In addition, each participant commonly used multiple
technologies as summarized by P12}:
\begin{quote}
    ``...I've got like Telegram, I've got Signal, I've got WhatsApp, I've got Kik, you name it, I've got it... 
    if there's something really large, that requires Dropbox... I'll do that''.
\end{quote}

With the exception of 5 participants, nearly all stored intimate content locally in their mobile device photo gallery,
laptop, or desktop computer. 
More than half 
intentionally stored their content in a cloud/file share platform.
The vast majority used multiple forms of storage: ``I have everything stored everywhere, I have it in my phone, I have it in my hard-drive. I have some parts on the cloud'' (P42). 

Participants decided to use technologies based on a variety of factors that we detail below. While we discuss each heuristic in isolation, most participants used several heuristics. For example, P17 explained: ``Encryption and security would be...number one. Number two, that is...convenience."

\label{subsect:decisions-safety}
\subsection{Convenience}
Around half of our participants shared intimate content using the same messaging service that they were already using to communicate with others.
Since participants often shared intimate content in the middle of a conversation, 
they emphasized the importance of conversation \textit{continuity}: ``...it's like at the heat of the moment... 
so to 
[choose another method and]
make it longer...
it just doesn't make any sense" (P43).
%
%
%
%
About a quarter also commented on the importance of \textit{usability} and highlighted different affordances that they valued such as
fast message delivery, desktop integration, clean user interfaces, and familiarity. Therefore, individuals that prioritize convenience must make do with any existing safety tools available on the platforms they use for sharing intimate content.

\subsection{Security}
%

\smallskip
\noindent\textbf{Anonymity.} 
Slightly less than a quarter of our participants sought \textit{anonymity} from  platforms and their recipients, particularly strangers, commonly as a means of protecting themselves from identification and other downstream harms in the event of recipient resharing (Section~\ref{subsect:threats-recipient}).
When meeting someone new from a dating app, P12 used a specific messaging app because ``you have a user name instead of it being connected to a phone number. [...] there's...a degree of arm's length-ness about it.'' 
Others, particularly sex workers, preferred platforms that allowed anonymity due to
concerns about platforms that make account recommendations using contacts, common connections, or shared location. 

\smallskip
\noindent\textbf{Security Feature Availability.}
Around a sixth of participants highlighted the importance having specific features available to defend against recipient resharing,
such as P4 who used
an app because it 
``prevents you from taking screenshots...
With people I don't know, and people... I have no trust in, I only send pictures through [this app].'' The features participants sought are discussed in greater depth in Section~\ref{sec:strategies}.
%

  \smallskip
\noindent\textbf{Data Control.}
It was important for under a quarter of participants to have \textit{data control and transparency} to prevent non-recipient threats (Section~\ref{subsect:threats-non-recipient}) such as data breaches and the unintentional discovery of intimate content.
They wanted to ensure that their content was not sent to a server they were unaware of (P2), that an app did not share images within its own parent ecosystem (P6), or the ability to select images an app had access to (P19).

 \smallskip
\noindent\textbf{Context Separation.} 
Three participants 
selected platforms to be exclusively used for intimate content to provide \textit{context separation} and prevent unintentional discovery (Section~\ref{subsect:threats-unintentional}). P1 explained: 
``I'm afraid that I [will] click the wrong person to send content...that's why I prefer to use \textit{[messaging app]}
which is an app that I don't use that much.''




\smallskip
\noindent\textbf{Reputation.}
The reputation of  technologies influenced a fifth of participants.
People leaned on folk models~\cite{wash_folk_2010} 
of ``secure'' or ``insecure'' platforms to make choices.
Some 
felt unsafe using certain platforms based on information they encountered about them, such as corporate data breaches~\cite{das2018breaking}, systems breaches~\cite{das2018breaking}, or general business practices like data sharing with third parties. 
A few were concerned that specific platforms had reputations tied to illicit behavior that would affect their recipients' perception of them. These ``social triggers''~\cite{das2019typology} led them to choose other platforms. P2 and P18 shared that they preferred popular platforms vs. a less well-known E2EE app that is ``marketed as being more secure'' (P2) since they ``don't know anybody that uses [\textit{that E2EE app}] unless they're using it for scandalous activities'' (P18). 
Conversely, others felt \textit{safer} using a sharing method because of a reputation associated with privacy. P31 used the same platform P2 was hesitant about because: 
``the ethos of the company...I know that is about privacy.'' We further discuss participant preferences around E2EE in Section~\ref{subsect:strategies-nonrecipients}
\subsection{Interpersonal}
\label{subsect:heuristics-interpersonal}
%

\smallskip
\noindent\textbf{Number of Active Users.}
About a sixth of our participants preferred not to use less popular platforms because they did not have enough recipients on that platform, which inhibited the \textit{convenience} of use.
For example, 
P9 tried an app that they felt was safer but 
``I only had two people there. So I deleted it.'' 
This can hinder adoption of platforms that participants may prefer for security reasons, like offering E2EE.

\smallskip
\noindent\textbf{Preference of Recipient.}
In the context of direct messaging, more than a quarter of participants were open to or deferred to (sometimes feeling like they had to) the technology preferred by their recipient. 
Particularly when the recipient prefers an app with specific safety features, 
some participants were willing to accommodate them.
Sex-working participants more commonly expressed that they needed to use technologies that felt more secure and convenient to their clients and also because in some cases, ``they're not willing [to] download like another app or like check another website'' (P32).
\subsection{Commercial}
\label{subsect:heuristics-commercial}
Finally, for around half of sex-working participants, profitability and hospitability to their work were key considerations.

\smallskip
\noindent\textbf{Profitability.}
A third of sex-working participants emphasized the \textit{profit potential} of a technology, such as a platform's ability to grow their client base, flexibility in setting rates (P20) and/or easy processes for receiving and transferring payments (P9, P24). 
Exemplifying the strong influence of profitability factors, P42 remarked, 
``I think [\textit{adult content platform}] is the worst in everything... But I use it a lot because it's the one that has the most people.''

\smallskip
\noindent\textbf{Hospitability.} 
A quarter of  sex-working participants valued a \textit{platform's hospitability toward sex work}. For some, it was
important for a platform to be made by sex workers and/or create an inclusive environment for queer people, such as P48 who found a platform "more comfortable" even though they thought it was complicated because, "...it's made by sex workers, [and] it's more like a gay, kinky, queer environment."

\section{Safety Maze: How People Try to Defend}
\label{sec:strategies}
All participants adopted at least one strategy but many relied on multiple. 
Since many prioritized convenience and the preference of a recipient, most participants leveraged any technological features available in the sharing platform and fill in significant gaps in availability of desired features (see Section~\ref{sec:gaps-recommendations-tech}) with interpersonal behavioral strategies and third-party tools. In this section, we describe participant strategies based on the threats they sought to defend against.

\subsection{Defending Against Recipient Resharing}
\label{subsect:strategies-recipients}
To prevent recipients from resharing intimate content, participants employed strategies across the timeline from first establishing trust with a recipient to after content had been shared. 

\subsubsection{Before Sharing}
 Prior to sharing intimate content,
 participants used interpersonal strategies to establish trust, communicate boundaries, and vet recipients: "with somebody new, I wouldn't necessarily send them something sexy until we've been on a couple of dates...
 until I knew that we were on the same page about privacy" (P29). With more established relationships, they commonly held an implicit expectation that their intimate content would be kept private. When sharing with friends, P31 explained, ``trust was so established that it didn't feel like you needed to say [not to reshare].'' In other settings, participants employed additional strategies that we now discuss.

\smallskip
\noindent\textbf{Vetting.}
Vetting a potential recipient of intimate content was described by over a third of 
participants, such as  speaking to recipients for some period of time prior to sharing content to assess whether they felt safe: ``I kind of establish a rapport and talk to people... [I'm looking for] just acknowledgement of the kind of precarity of the situation" (P12).
 A few participants relied on their intuition even without a conversation, such as P51 who shared intimate images through a dating app and described developing 
``a sixth sense'' that they used to ``make a judgement based on various features on the profiles'' 
to establish if the recipient 
would be unlikely to reshare content.
Many sex workers relied on payment for vetting. If the worker had payment details, they could take them to be
an implicit understanding that the content was for private use only. 

\smallskip
\noindent\textbf{Rule Setting.}
About half of participants explicitly asked recipients not to share or store intimate content.
``...I asked them not to screenshot anything, just to [keep] it in chat. And that was kind of the ground rule'' (P11). Sex workers were more likely to set explicit rules. P35 shared: ``before I send images, I always just say that this image is between myself and yourself...
No one else is paying for this except you.'' 
Recreational sexters mentioned that they would circumstantially ask for content not to be shared such as if they had friends in common (P4) or if there were identifying features in the content (P14). 
In addition, a few participants stated direct consequences if content were to be shared. 
For instance, P30 said, 
``I would just tell them, if you ever shared, I will come up to you, we will go to the police'' whereas P34 sent reminders to clients about potential consequences:
``I send them the penal law articles regarding sharing/publishing porn content.''

On the other hand, a few participants were open to recipients sharing intimate content if they asked permission first or in specific contexts: ``I have on occasion, given people permission to share it with other people, especially in the context of [initiating group sex]'' (P12).

\subsubsection{While Sending}
When sharing intimate content, participants were highly cognizant of the risks of NDII and the limitations of existing technical defenses: 
``yeah, I can secure the communication, I can secure the storage where the photo is in, but I don't think I can prevent anyone from actually getting pictures'' (P26). 

Rather than focusing on complete prevention, participants used various strategies because 
``any of these measures that I'm talking about [are]  a form of harm reduction'' (P46).
They therefore sought out technical features as a means of creating friction and used a combination of tactics to create accountability in event of NDII. Lastly, they sought to reduce the reshare value of their content so that it is less likely to be of interest to other audiences and proliferate further.

\smallskip \noindent
\textbf{Friction.} 
To create friction and reduce opportunities for recipient resharing, participants used \textit{expiring messages} that allow a sender to set a message expiration time. This was widely used by three-fifths of participants (mainly recreational) and many used it situationally with strangers or in higher-risk contexts:
``when I go to [\textit{country}], which is a more dangerous place for homosexuals, I use this expiring pic feature'' (P4). 

When available, a few participants also made use of \textit{screenshot prevention} in which screenshots result in a blank screen or deterrence messaging from a platform. P15 described using a platform because, ``you couldn't also take like screenshot, like it wasn't allowed like like how it is [on a banking app]." 
Since this feature was not readily available in most platforms used for sharing content (see Appendix~\ref{appendix:feature-availability}), it was also commonly mentioned by others as a feature they wished more platforms would implement it as a means of increasing friction, despite many noting that it could be bypassed (e.g. capturing content through another device).

\smallskip 
\noindent
\textbf{Accountability.}
Participants also used available safety features to reinforce boundaries and create mechanisms of accountability. For example, a third of participants used \textit{screenshot notifications} (particularly recreational ones) primarily to calibrate trust and ``make the confrontations and the discussions that you need to do to be safe'' (P10). P2 added that, ``[a screenshot notification] empowers [people] to make decisions about whether to continue sharing messages." This was exemplified by a scenario in which P5 asked a recipient to delete an image after receiving a screenshot notification.

\textit{Expiring messages} were also used in this context to calibrate the sensitivity of the content (e.g. sharing more explicit content) and implicitly communicate boundaries around resharing: ``it’s a clear, unambiguous sort of caption that says I would prefer if you didn't hold on to this'' (P2).

Similarly, participants added watermarks and messaging on content to  imply that it should be kept private and/or to trace leaked content back to its source. For example, P31 used watermarks on commercial content and also wrote the name of their recipient onto an image in their personal life to `put the accountability
on the person...
you sort of implicate them into it more.'' P41 wrote a recipient's username on content as a way of later identifying who shared it.
Participants also used watermarks to proactively prepare for future removal of nonconsensually shared content since ``when you report on social platforms, you have proof of ownership'' (P1). 

Lastly, a few participants shared in contexts where there was a mutual exchange of intimate content: ``if someone has also shared content of themselves, it kind of makes me feel better to also share something" (P47). 
For P26 who shared \textit{and received} content in a network of strangers (see Section~\ref{sec:use:case}), they "expect them to not share [my intimate content] because they're in the same situation as me."

\smallskip
\noindent
\textbf{Lowering Content Value.}
To prevent resharing was, participant made content less desirable to others aside from intended recipients. 
Avoiding identifying features (Section~\ref{subsect:strategies-identification}) not only gave participants more anonymity, but also made intimate content less valuable and possibly less likely to be reshared.  Personalizing content (e.g. writing a recipient's name) served a similar purpose.
Lastly watermarks, particularly in commercial contexts, because 
``if your content's watermarked, then other platforms won't allow it to be uploaded'' (P35).

\subsubsection{After Sharing}
After content had been shared, participants used strategies to manage and revoke access to reflect changing dynamics of trust in a sharing context.

\smallskip \noindent \textbf{Deletion Request.}
Although in some contexts participants consented to their recipients storing their intimate content, almost a third had made an explicit deletion request at some point. 
Sex-working participants made these requests in their personal lives and with clients. With clients, P20 shared that this occurred ``when I sent them something that I later reconsidered... 
some things that I'm not comfortable with.'' However, P1 pointed out that this may be difficult after a transaction, ``I even offered to like refund but no, [the client] didn't [delete the content].''
Across contexts, participants expressed that they were unsure if content was ultimately deleted, such as P50 who described an ``emotionally charged'' period where they were not sure if their former partner had deleted their intimate images after a breakup.

\smallskip
\noindent\textbf{Retroactive Unsend.}
Around a quarter of participants manually deleted intimate content from messaging  histories such as through using an ``unsend'' feature.
P32 described waiting for a client to view an image. After they receive a read receipt, they delete the image from their 
conversation history. P5, who shared with strangers online, added that, ``So sometimes if I feel like you know, a person's acting fishy, then I either delete the messages or delete the photos.''

\subsection{Defending Against Identification.}
\label{subsect:strategies-identification}
Participants proactively used strategies to prevent being identified in the event of NDII. 
This may be to create plausible deniability (e.g., against blackmail) and to protect against additional harm (e.g., doxxing, harassment). 
In many cases, participants employed these strategies with the understanding that once content was shared, it was out of their control. ``I know that I don't have full control over where these images will go, if people can trace them back to me" (P26).

\smallskip
\noindent\textbf{Avoiding Identifying Features.}
Half of participants avoided capturing identifying visual information when creating content or removed such information after creation. 
Most commonly, they chose not to include their face or tattoos 
by using photo editing tools (to crop, blur, or cover them) or bandages/clothing
such as P37:
``Because I have big tattoos on my legs, I wear high knee high socks.''
Participants also voiced concerns about being identified through the background in their image or video, 
``the language of a shampoo bottle, street signs/public places… are carefully edited out'' (P33).
Although more than a third of participants used photo editing tools (either in-app or third-party),
there were a few security concerns
since they would ``have to upload my intimate images to that app as well. So it feels extremely unsafe'' (P23).
Aside from using photo editing tools,
participants also chose neutral backgrounds, removed identifying items in their background prior to taking content, and/or frequently changed the location in which content was created.


\smallskip
\noindent\textbf{Metadata Removal.} Six participants used metadata removal to eliminate extra information that is added when an image is created (e.g., location, date/time of creation). However, they generally expressed that they would prefer for it to be automatically done by a platform: ``[it] almost feels like that should be the default when you share any image anyway'' (P6).


\subsection{Defending Against Non-Recipients}
\label{subsect:strategies-nonrecipients}
Along with concerns about recipient resharing, participants sought to defend against those who may access or view their intimate content through data breaches, hacking, and/or unintentional discovery (Section~\ref{subsect:threats-non-recipient}).

\smallskip
\noindent\textbf{Storage Strategies.}
To defend against storage-related 
(e.g., hacking, unintentional discovery, device theft), around half of our participants proactively deleted their own stored content:
``my [\textit{cloud storage}] will send me like, memories of [intimate content]...
[when I'm] 
trying to sip my tea... That's where I'll be like, yeah, get out of here" (P46).
A quarter were mindful about their authentication practices---choosing to use strong passwords, 2FA, biometric authentication, or password managers. Finally, a quarter of participants stored content in local storage (or disabled automatic backups) while another quarter used a hidden folder, which moves selected photos from a photo gallery into a designated folder (e.g., Apple \cite{apple_hide_nodate}, Samsung\cite{samsung_what_nodate})
to prevent unintentional discovery.

\smallskip
\noindent\textbf{Consent.}
Many participants mentioned the importance of consent prior to sharing intimate content. A few specifically
mentioned asking for consent before sharing intimate content in order to check if a recipient was alone or in a private area to prevent shoulder surfing, ``I'll always ask first and wait for their response before I send something because you never know what people are doing'' (P29).

\smallskip 
\noindent
\textbf{Masking Content.} Another strategy a few participants used to protect themselves against shoulder surfers was to mask it upon sending it. For the two participants who used this feature, a recipient would receive a blurred image that they would then have to click on to reveal the contents. This would signal that the content was sensitive and should be viewed in a private location. Though this feature was not commonly mentioned, it is also not readily available in most platforms (see Section~\ref{sec:gaps-recommendations-tech}).

\smallskip
\noindent\textbf{End-to-End Encryption.}
End-to-end encryption (E2EE) ensures that only a sender and their intended recipient(s) can view a message.
Once an E2EE message has been delivered, it carries similar risks to unencrypted messages: the recipient can still
store or share message contents (some platforms may auto-download images onto devices). Over a quarter of participants actively used E2EE or considered it to be useful (excluding those who happened to use platforms with E2EE but did not mention it). For a few, E2EE alleviated their concerns about unwanted access by a platform or government entity: 
``There are basically two possible attackers. Like either the person I'm sending this to or someone from the outside trying to break into systems...
I mostly care about someone from the outside able to steal pictures'' (P26). 
P10 added that they valued E2EE because, ``if there's anything I don't want, it's the government knowing details of my personal life and seeing my tits.''
On the other hand, a sixth of participants did not trust E2EE, believing that it could be broken if needed:
``If someone wants the data, they will get their hands on it anyway. And if that someone is the police, they will just force companies to give that to them. No matter how well encrypted it is'' (P50).

\subsection{Comparing Participant Groups}
\label{subsect:comparisons}


%
%


In our work, we purposefully included the perspectives of victim-survivors and sex workers. Among our 52 participants, 22 were victim-survivors (8 recreational, 14 commercial).

We encountered narratives in which experiencing harm had increased fears of NDII: that, "I'm just really paranoid... and there's nothing I can really do about it... Especially after it's already happened" (P35). 
We also observed that some victim-survivors implemented additional strategies after experiencing harm. 
As P21 explained, ``What happened to me was forever... And so I was more knowledgeable about what was going on, about the privacy of these pictures.'' P43 shared, "future photos... were really less revealing than they used to... you cannot recognise me." 
Otherwise, we did not observe notable differences in practices.
Since we interviewed victim-survivors who continued sharing intimate content, many expressed that they resolved not to let shame and stigma stop them (and for some sex workers, they were unable to due to financial reasons). For example, P18 shared:
\begin{quote}
    "I'm having fun. I'm an adult...
I don't want to allow shame or stigma to prevent me from living my life and exploring my sexuality."

\end{quote}


When comparing commercial and recreational participants, commercial participants shared content in all the same contexts along with specific platforms to sell content and/or advertise services. However, sex workers were more constrained by the preferences of their clients when selecting a sharing method. Commercial participants may also prefer sex work specific platforms since they may otherwise perceive or experience hostility.
We also observed that sex workers were less worried about unintentional discovery, either since the people around them already knew they were sex workers or because they employed more strategies to prevent this.

Generally, we observed that while recreational participants adopted many strategies, sex-working participants adopted more (in particular, 
watermarking, removing metadata, and using safe storage methods). They were also more likely to set explicit boundaries around sharing or storing.
This greater protective action may be due to both  differences in threat models (e.g. greater personal/physical risk, stigma)
as well as better advice and educational support (see Section~\ref{subsect:advice}).

\section{Gaps in Defenses \& Recommendations}
\label{sec:gaps-recommendations}

While work on secure technology to address IBSA is limited,
we are not the first to study this topic (see Section~\ref{sec:related}). Our results build on prior work, such as Geeng et al. ~\cite{christine_geeng_usable_2020}. Taking our results together, we see that participants use a range of technologies spanning different contexts and have varied threats from NDII to accidental forms of content discovery. We also observe that users leverage technological features when available. In both our findings, the most common feature used was expiring messages, perhaps due to its wider availability across platforms and popularized use with Snapchat. This supports our recommendation for wider platform adoption of safety features (discussed below).
Further, our work in conjunction with Waldman \cite{waldman_law_2019}, demonstrates that individuals often rely on interpersonal strategies to supplement (or in lieu of sufficient) content-based strategies. We echo the prior~\cite{christine_geeng_usable_2020} suggestion for platforms to investigate design patterns that assist users in negotiating boundaries like consent-based design~\cite{zytko2022consent} and the need for platforms to design for intimate content as a use case, rather than as misuse.
In addition, our findings reveal new gaps and opportunities for technological improvements alongside research in order to create more robust solutions against NDII.

\subsection{Technological Recommendations}
\label{sec:recommendations:platforms}
\label{sec:gaps-recommendations-tech}
Our 52 participants used over 40 platforms for intimate content. Thus, 
any platform that enables the storing or sharing of images is likely being used for intimate content. Platforms must re-evaluate the sensitivity with which image and video data is treated and how it is secured.
We detail high-level user needs below. 
We emphasize that such features should be flexible, allowing people to adjust to different interpersonal strategies and contexts (Section~\ref{sec:use:case}). In conjunction, we also urge for additional research to understand the limitations of these features (discussed further in Section~\ref{subsect:gaps-research}).

\smallskip
\noindent\textbf{Availability \& Visibility of Existing Features.}
Many participants felt that they could or should be doing more to protect themselves but 
the features they wanted were not always available.
 Further, using protective strategies is  time consuming, as 
 P33 noted, 
 ``I spend more time scrubbing personal info off my pictures, putting them in private folders, etc., than actually taking said pictures. The learning curve is steep.''
Participants also sometimes forgot to use features because of the burden of using them.
  As a result, a few expressed a sense of futility: ``I do what I can... [but] I just sometimes think it's pointless'' (P23). P9 added, 
  ``It's like I gave up my security...
  everything I have read, it seems too complicated."

Participants wanted a full suite of safety features in any platform that allows photo/video storage and sharing. 
Desired features included: 
hidden folders
, password protections
, automatic watermarking
, retroactive delete/unsend
, expiring messages, masking on send, options to disable automatic download or backup of content sent in a messaging app, and screenshot notification.
As P12 described, ``If you could take the anonymity of [\textit{platform}], the disappearing thing that you get from [\textit{another platform}]. And this security that you get from [\textit{E2EE platform}], that would be quite cool.'' 

In response to these narratives, we conducted a systematic analysis of 8 popular platforms (see Appendix~\ref{appendix:feature-availability}) of the availability of safety features participants desired.
We found that none of them fully implement all of the features (or features were only available in a special mode/for paid users).
We emphasize that wider adoption of features is needed, particularly in popularly used platforms, to protect the majority of sexters who prioritize convenience or defer to their recipient's choice.


\smallskip
\noindent
\textbf{Content-Level Data Control \& Presets.} 
Participants wanted variations of existing features for data control that better reflect how they navigate trust and to help separate contexts.

Specifically, participants desired modifications and/or expansions of existing features to operate on a per-message (or per content) basis: 
being able to make an specific message expire; allowing screenshots only a single piece of content; or applying a shared password that would be needed to open a specific piece of content\footnote{While WhatsApp recently released the ability to lock individual conversations, to our knowledge, no technologies allow password protection of a single message \cite{whatsapp_whatsapp_2023}. Prior work~\cite{hasinoff_sexting_2014} offers preliminary quantitative evidence that a majority of young adult sexters would use such a feature.} ``to make sure that no one else except that certain someone opens it'' (P28). To prevent shoulder surfing, P23 suggested clicking a button or a sticker to reveal intimate content, rather than displaying it in a message window.\footnote{Confide (\url{https://getconfide.com/}) offers such functionality; no participants in our study had heard of the application.}
Lastly, they wanted the guarantee that when they delete content, it is deleted for recipients and that platforms no longer store the content, 
``you can delete everything [\textit{in messaging app}] for yourself and the other person...
it's not stored anywhere, it's not backed up or anything'' (P35).

However, manipulating features based on the content or interaction is cumbersome.
Thus, participants suggested offering a preset ``intimate picture mode'' (P30) in messaging apps, or several modes with preset settings appropriate for interactions with different levels of trust.
This suggestion aligns with best practice from academic privacy literature \cite{knijnenburg_privacy_2017, wisniewski_give_2015}, which recommends privacy presets based on user contexts or personas. 
Since increasing the granularity of control may make platforms more difficult to use, additional research is needed to balance this with maintaining ease of use.

\smallskip
\noindent
\textbf{Support Manual Strategies.} 
Participants wanted more technological support for the strategies they already used by making features easier to use and introducing new ones to complement interpersonal strategies. 
For example, participants wanted to \textit{automate existing manual content-level strategies} (e.g., automatic background removal or face blurring in stand-alone or in-messaging-app photo editing tools or automatic metadata removal when content is captured). 
Participants also suggested various \textit{design patterns to support interpersonal strategies} such as 
%
behavioral nudging: ``a reminder that if you share this image without the consent, you could be imprisoned for time, maybe an alert to a service to contact the the police or psychologist'' (P30) 
or a guided process of mutual consent to turn on storage of content sent in messages (P12).



\smallskip
\noindent
\textbf{Secure Storage.}
Participants wanted more and better options for secure storage. Six participants 
wanted expanded options for secure local storage or encrypted storage. ``If there would be an app or method intended for intimate image storage only, I would totally use it to prevent any security issues...
something that isn't connected to the internet'' (P14). Some emphasized wanting to ensure locally stored intimate content was not accessible to other apps.
A few participants suggested building a designated app for storing intimate content where content would be stored locally in an encrypted format and that they could then selectively share content from. Although this sounds similar to vault apps,\footnote{Vault apps are designated apps for storing sensitive content such as images or documents but may offer varying levels of security \cite{ruffin_casing_2022}. They are often disguised as other apps to avoid discovery when device sharing. See~\cite{vault} for a more in-depth discussion on the use of vault apps for intimate content storage.} only one participant had used one. Similarly, a few others wanted a mechanism for sharing a photo album that they could easily revoke access to: ``I have never seen any app...that would allow you to like, you know, sort of create a photo album inside of your conversation'' (P7). This suggestion is similar to Grindr's Album feature \cite{grindr_albums_nodate}, which multiple participants wished more platforms would adopt.

\smallskip
\noindent\textbf{Increase Friction.}
As previously noted in Section~\ref{subsect:strategies-recipients}, participants desired additional features for increasing friction since,
``No matter how much features people put into safety, there's always always always going to be a risk of it...[but] as long as the features and the way you're doing it has the minimum level of safety that will [stop] most people'' (P16).
Such features included download notifications (for when recipients save content and/or designating content for which download should be offered) or screenshot deterrence. While these features are not without risk (discussed further in Section~\ref{subsect:gaps-research}), allowing users to disable auto-downloading of images/videos in messaging platforms or remove the download option for designated content can be meaningful changes. 

\smallskip
\noindent\textbf{Improve Transparency.} 
For each feature we investigated, some participants expressed valid concerns while others had misconceptions.
Particularly with E2EE, misunderstandings about the scope of protection and inconsistent implementations 
led to wholesale distrust, which is consistent with findings from prior work on E2EE (Section~\ref{sec:related}). Users also wanted clarity on how data persisted on platform servers when using expiring messages or other mechanisms of deletion. For example, P51 expressed the concern that, ``even if [a recipient] deletes [stored intimate content], you delete it from your phone or wherever, it will be stored somewhere in some server.'' When deleting content, P10 added, ``in the ideal sense... I want to be able to send the image and then to have it be wiped off the face of the planet and off the data server that it's in.''
There is need for more transparency around protections offered in this high-risk context and new mechanisms for guaranteeing/verifying deletion.

\subsection{Research Directions}
\label{subsect:gaps-research}
\smallskip
\noindent 
\textbf{Concerns around Safety Features.}
Though technical features (e.g. screenshot notifications, expiring messages, watermarks) were widely used, this was often caveated with expectations that these protections could be bypassed. 
While watermarks served as protection against content theft and created accountability, many acknowledged that they could altered or 
``used against people like, oh, it's the same girl in all these photos, collect them and put them in a folder together.'' 

For screenshot notifications, screenshot prevention, and mechanisms that provided ephemerality, participants acknowledged that content could still be captured by another device. Despite these caveats, participants desired these features as means of increasing friction. However, it is important that the limitations of these features are clearly communicated to users so as to not give false expectations around security.


In addition to these concerns, some participants feared that such features would incentivize the behavior they sought to prevent against. For example, P31 was concerned that expiring messages
``sort of invite people trying to screenshot...
[because] it almost creates this environment of, oh, this is secret, which sort of invites people to be like, Oh, I'm going to try and hold on to it'' (P31). 
Further investigation is needed to understand if there are any contexts in which these features may promote the capturing and storage of intimate content.

\smallskip
\noindent
\textbf{Balancing Content Control with Harm Documentation.}
Participants wanted message ephemerality (e.g., expiring messages, unsending a message) for protection, especially in high-risk contexts. However, in the event of NDII, it may be useful (or necessary) to have a record of the sender, recipient, and content of messages for pursuing legal action or content removal.
In the context of unsending messages, P38 suggested removing the content only for the recipient. 
This and other design ideas should be explored to resolve the desire for ephemerality while allowing senders to maintain documentation in case of harm. However, it is important that such designs support victim-survivors while not introducing potential mechanisms for censorship and/or other privacy violations.


\smallskip
\noindent
\textbf{Discovery of NDII.} Many participants wanted more technologies for identifying nonconsensual sharing.
P33 described this as ``an image search built in that would notify us when our content is posted elsewhere.''
More specifically, P47 wanted ``a way for the actual content shared through the app to be identifiable through some kind of hash signature.'' 
We broadly referred to this functionality as ``digital fingerprinting''~\cite{wagner1983fingerprinting} with participants. 
Only three participants had heard of this and only one used something related: a facial recognition tool.
 
 After being provided a high-level description (see Appendix \ref{appendix:questions}), participants had mixed reactions of both interest and skepticism. 
 A few  
 expressed enthusiasm about this functionality enabling them to more easily find NDII content and perpetrators 
 but others had reservations 
 around the security of fingerprint storage and whether the fingerprint itself could be used to identify or otherwise violate the image owner's privacy: 
 ``what if it could be traced back to me?'' (P10). 
Meanwhile, a few participants 
commented that identifying the content alone would not be useful since barriers exist to removing nonconsensually shared content on many platforms. 




Organizations like StopNCII~\cite{stopncii_how_2023} and AmIInPorn \cite{am_i_in_porn_faq_2023}  aim to offer such functionality but lack public implementation details, making it difficult to assess any risks. 
These services may rely on perceptual hashing, which produces the same hash for \textit{similar} content, or facial recognition. Perceptual hashing has previously been vulnerable to attacks that recover partial information about the original image~\cite{sarah_scheffler_sok_2023}, justifying participant concerns. Both approaches can have high false positive rates~\cite{sarah_scheffler_sok_2023,buolamwini_gender_2018}, which may mistakenly alert individuals that an image was shared nonconsensually, causing distress.


More work is needed to explore technical mechanisms for tracking and/or identifying NDII, specifically to: investigate security/utility tradeoffs, establish effective transparency mechanisms, 
establish design paradigms for integration into platforms, and further investigate user concerns. Our recommendation to further investigate such technologies
should not be taken as an endorsement of these methods, especially when applied in other contexts~\cite{carmela_troncoso_joint_2023, danielle_keats_citron_criminalizing_2014, sarah_scheffler_sok_2023, seny_kamara_outside_2021,marks2021can}.

\smallskip
\noindent 
\textbf{Insufficient Education.}
\label{subsect:advice}
Many recreational sexters had not ever encountered advice for sharing intimate content nor had they sought it out,
because, ``it's not something that's talked about too often'' (P3). As a result, they stumbled across safety features and navigated mitigation strategies on their own.

Those who had not experienced NDII, were generally unfamiliar with post-harm mitigation strategies and resources. We asked non-victim-survivors for advice they would give to a hypothetical friend who had experienced NDII\footnote{Victim-survivors were not asked this question and instead asked about their experiences with NDII, which we plan to report on in future work.}.
The most common response was to go to the police and/or seek legal action, which may be inaccessible options (see Section \ref{sec:related}). 
A fifth mentioned reporting content to a platform. Only one to two participants mentioned looking for help online and seeking help from an organization.
Many also admitted that they were unsure what to do in this situation.

In contrast, a majority of sex-working participants spoke about the vast resources available to them
through online sex worker communities. They were able to access safety content made by sex workers, ask questions freely, and receive support. This resulted in sex-working participants  implementing more strategies such as watermarking, removing identifying features, safe storage, and/or metadata removal. 

Broadly, there is a need for visible and
accessible resources for people to learn how to protect themselves against NDII. 
Presently, the most common advice participants received from society, media, and others was not to share any intimate content. This was viewed as impractical, stigmatizing, and responsible for creating a culture of victim blaming because:
\begin{quote}
``everyone does it... it's a part of how we sort of sexually express ourselves, or make money...
If something does happen, and it does go badly, then you feel like, \textquotesingle Oh, well, everyone's just saying, I shouldn't have done it.\textquotesingle 
So somehow, it's my fault even though it's obviously not''
(P31). 
\end{quote}

Research such as ours can facilitate the creation of accessible advice guides for end users, particularly with the inclusion of resources from sex workers (since they practice more strategies and have experience educating and disseminating safety practices). Future work should look toward establishing such resources, evaluating their efficacy, and work on making resources more discoverable since end users may not be actively searching for advice due to the reasons discussed.
\subsection{Concluding Discussion}



\label{sec:conclusion}
We conclude by observing the ways in which societal stigma is encoded by technology's lack of protection against NDII and the broader cultural changes needed to make progress toward reducing NDII.

\smallskip
\noindent
\textbf{Technology \& Victim-Blaming.}
Participants repeatedly mentioned themes of personal responsibility. They acknowledged the risk ``inherent'' in sharing intimate content, expressed fear of exposure and desire for greater control and autonomy. They expressed a matter-of-fact recognition that responsibility for their safety had been displaced onto them and that platforms were notably absent as ``capable guardians to intervene''~\cite{henry_image-based_2022}. 
Left to navigate safety on their own, participants must shoulder the burden of preventing NDII and other threats (see Section~\ref{sec:strategies}). The volume and stringency of security behaviors that our participants engaged in extends beyond the standard level of responsibility any individual sharing online content should be reasonably expected to assume. Despite the effort they put into security, participants critiqued themselves for not implementing enough behaviors.  
%
\textbf{We conceptualize this as a form of victim blaming enacted by the absence of protective technological design}, which communicates the belief that a victim of sexual violence is responsible for (preventing) the consequences.
  As a result, it is critical to \textbf{reallocate responsibility from individuals to platforms and technologists.} 
We need to reconsider how much personal responsibility individuals must assume when engaging in sharing intimate content and how much of that responsibility can be reallocated or alleviated by implementing technological features. When possible, we must prioritize making technology that works in favor of individuals’ autonomy, and safety, and implement features that relieve the burden from users. Individuals should not have to pick between avoiding abuse and engaging in recreation or labor. 

\smallskip
\noindent\textbf{All of Us: Reduce Stigma, Bring  Consent.}
Alongside platform changes and state-of-the-art technologies, participants expressed the desire for societal change. They vocalized wanting educational change that focuses on harm reduction and a shift from punitive ways of addressing harms. 
They envisioned a future where asking consent to share is the norm. ``We should get into a place of cultural [understanding] where if I send you a picture of my dick, and you show it to someone else, they're like, did they fucking consent to that?'' (P38).

As some participants have had to address their own shame, they found comfort and
``feel so much safer about my nudes when I think about how many people send nudes'' (P49). And with how common it is, participants hoped for a future where there is no longer stigma attached to sharing intimate content:
\begin{quote}
    ``I really hope that culture starts to change... 
    enough people are sending sexually explicit messages that our next generation of politicians will all have nudes in the world. So hopefully it won't be a scandal. People will just be like, hey, nice tits, great'' (P18).
\end{quote}

\section*{Acknowledgments}
We thank our participants for sharing their experiences with us. This work was supported in part by the European Sex Workers Rights Alliance, the Max Planck Institute for Software Systems, and the U.S. National Science Foundation under award CNS-2206950. Lucy Qin was additionally supported by the NSF Graduate Research Fellowship.
We thank Allison McDonald for feedback on an earlier draft of this work. 


{\footnotesize
\bibliographystyle{plain}
\bibliography{main}
}

\appendix
\section{Feature Availability Across Platforms}
\label{appendix:feature-availability}
We conducted an analysis of feature availability in 8 popular platforms in November 2023. We selected features based on what participants reported using and what they shared that they wanted in a platform for sharing intimate content. 
The results are displayed in Table~\ref{table:feature-availability}. We designated a platform as providing a feature if it was available in some variant. This includes when it is only available in a specific mode, such as "vanish mode" in Instagram or a hidden/locked chat in Whatsapp. To assess the availability of features, we relied on published documentation and user guides from the platforms themselves. In some cases, we directly downloaded and accessed the platforms we surveyed to confirm the availability/unavailability of a feature when needed. 

\newcolumntype{R}[2]{%
    >{\adjustbox{angle=#1,lap=\width-(#2)}\bgroup}%
    l%
    <{\egroup}%
}

\newcommand*\rot{\multicolumn{1}{R{90}{1em}}}

\makeatletter
\def\adl@drawiv#1#2#3{%
        \hskip.5\tabcolsep
        \xleaders#3{#2.5\@tempdimb #1{1}#2.5\@tempdimb}%
                #2\z@ plus1fil minus1fil\relax
        \hskip.5\tabcolsep}
\newcommand{\cdashlinelr}[1]{%
  \noalign{\vskip\aboverulesep
           \global\let\@dashdrawstore\adl@draw
           \global\let\adl@draw\adl@drawiv}
  \cdashline{#1}
  \noalign{\global\let\adl@draw\@dashdrawstore
           \vskip\belowrulesep}}
\makeatother
\setlength{\dashlinedash}{0.5\dashlinedash}

\NiceMatrixOptions
{
    custom-line =
    {letter = I ,
    tikz = {line width=0.25pt, gray!40 },
    width=0.25pt   
    },  
}

\newcommand{\SurveyTrue}{{\color{black}\CIRCLE}}
\newcommand{\SurveyPartial}{{\color{black!55}\LEFTcircle}}
\newcommand{\SurveyFalse}{{\color{black!30}\Circle}}


\begin{table} 
\centering
\footnotesize
    \begin{tabular}{l|c|c|c|c|c|c|c|c|c|}

      & \rotatebox[origin=c]{90}{{\bf Apple iMessage}} & \bf \rotatebox[origin=c]{90}{{  \bf Discord}} & \bf \rotatebox[origin=c]{90}{{  \bf Google Messages}}  & \bf \rotatebox[origin=c]{90}{{  \bf Grindr}} & \bf \rotatebox[origin=c]{90}{{  \bf Instagram DM}}  & \bf \rotatebox[origin=c]{90}{{  \bf Signal}}  & \bf \rotatebox[origin=c]{90}{{  \bf Snapchat}} & \bf \rotatebox[origin=c]{90}{{  \bf WhatsApp}}       \\ 
         \\
        \toprule
          Screenshot Prevention & \SurveyFalse & \SurveyFalse  & \SurveyFalse  & \SurveyTrue  & \SurveyFalse  & \SurveyFalse  & \SurveyFalse  & \SurveyTrue \\
        
       Download Prevention & \SurveyFalse & \SurveyFalse  & \SurveyFalse  & \SurveyTrue   & \SurveyTrue & \SurveyTrue  & \SurveyTrue  & \SurveyTrue \\
       
       Screenshot Notification & \SurveyFalse & \SurveyFalse    & \SurveyFalse  & \SurveyTrue  & \SurveyFalse & \SurveyFalse  & \SurveyTrue  & \SurveyFalse \\

       Watermarking & \SurveyPartial & \SurveyFalse  & \SurveyFalse  & \SurveyFalse  & \SurveyFalse   & \SurveyPartial  & \SurveyPartial  & \SurveyPartial \\

        Mask on Send & \SurveyFalse & \SurveyPartial  & \SurveyTrue  & \SurveyTrue  & \SurveyFalse & \SurveyTrue  & \SurveyTrue  & \SurveyTrue \\

       Metadata Removal & \SurveyTrue & \SurveyFalse  & \SurveyTrue  & \SurveyTrue  & \SurveyTrue   & \SurveyTrue  & \SurveyTrue  & \SurveyTrue \\

        Blur Identifying Features & \SurveyPartial & \SurveyFalse  & \SurveyFalse    &  \SurveyFalse  & \SurveyFalse &  \SurveyTrue  & \SurveyPartial  & \SurveyPartial \\

    Message Unsend & \SurveyFalse & \SurveyTrue  & \SurveyTrue  & \SurveyPartial  &\SurveyTrue & \SurveyTrue  & \SurveyTrue  & \SurveyTrue \\

    Expiring Message & \SurveyFalse & \SurveyFalse  & \SurveyPartial  & \SurveyPartial & \SurveyTrue &  \SurveyTrue  & \SurveyTrue  & \SurveyTrue \\
        \bottomrule
    \end{tabular}

    \caption{Survey of feature availability in popular platforms used for sharing intimate content. \SurveyTrue $ $ represents the availability of some variant of a feature, although this may be isolated to a specific mode of the platform, such as ``vanish mode'' in Instagram. \SurveyPartial $ $ represents the partial availability of a feature due to it being available only to paid users or through using a different feature for an unintended purpose, such as creating using decorative stickers in photo editors to cover identifying features. \SurveyFalse $ $ represents the unavailability of a feature in any form.}
     \label{table:feature-availability}

\end{table}

\section{Methods (Continued)}
\subsection{Recruitment Graphics}
\label{subsect:appendix-recruitment}
The following recruitment graphics were used to recruit recreational participants (Figure 1) and commercial participants (Figure 2).

\subsection{Privacy and Anonymity.}
\label{subsect:appendix-privacy}
To protect the privacy and anonymity of the participants, 
we used end-to-end encrypted platforms for the interviews, provided advice on how to sign up with an encrypted (and anonymous) email address if preferred, and used Calendly to schedule interviews, which did not require any legal information or login. The payment options could be anonymously received as well. Interviews were end-to-end encrypted and conducted either using Webex or Zoom. 
Participants were given the option of conducting the interview as a video call, audio call, or text chat. 
Sex-working participants were given the option of being interviewed by a current or former sex worker. 

\begin{figure}[ht]
\label{fig:recruitment-graphic-recreational}
\centering
\includegraphics[width=0.4\textwidth]{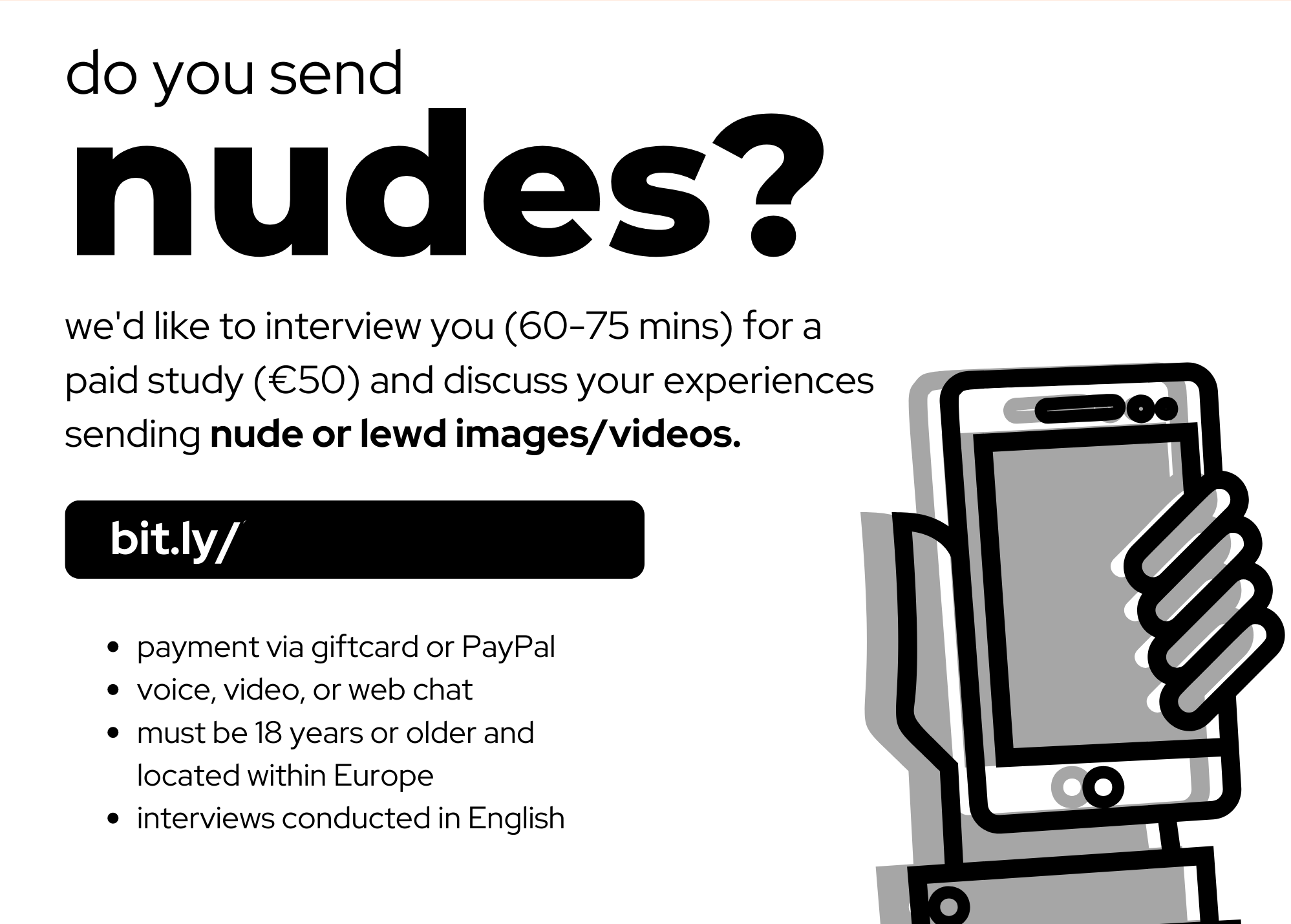}
\caption{Recruitment graphic for recreational sexters}
\end{figure}

\begin{figure}[ht]
\label{fig:recruitment-graphic-commercial}
\centering
\includegraphics[width=0.4\textwidth]{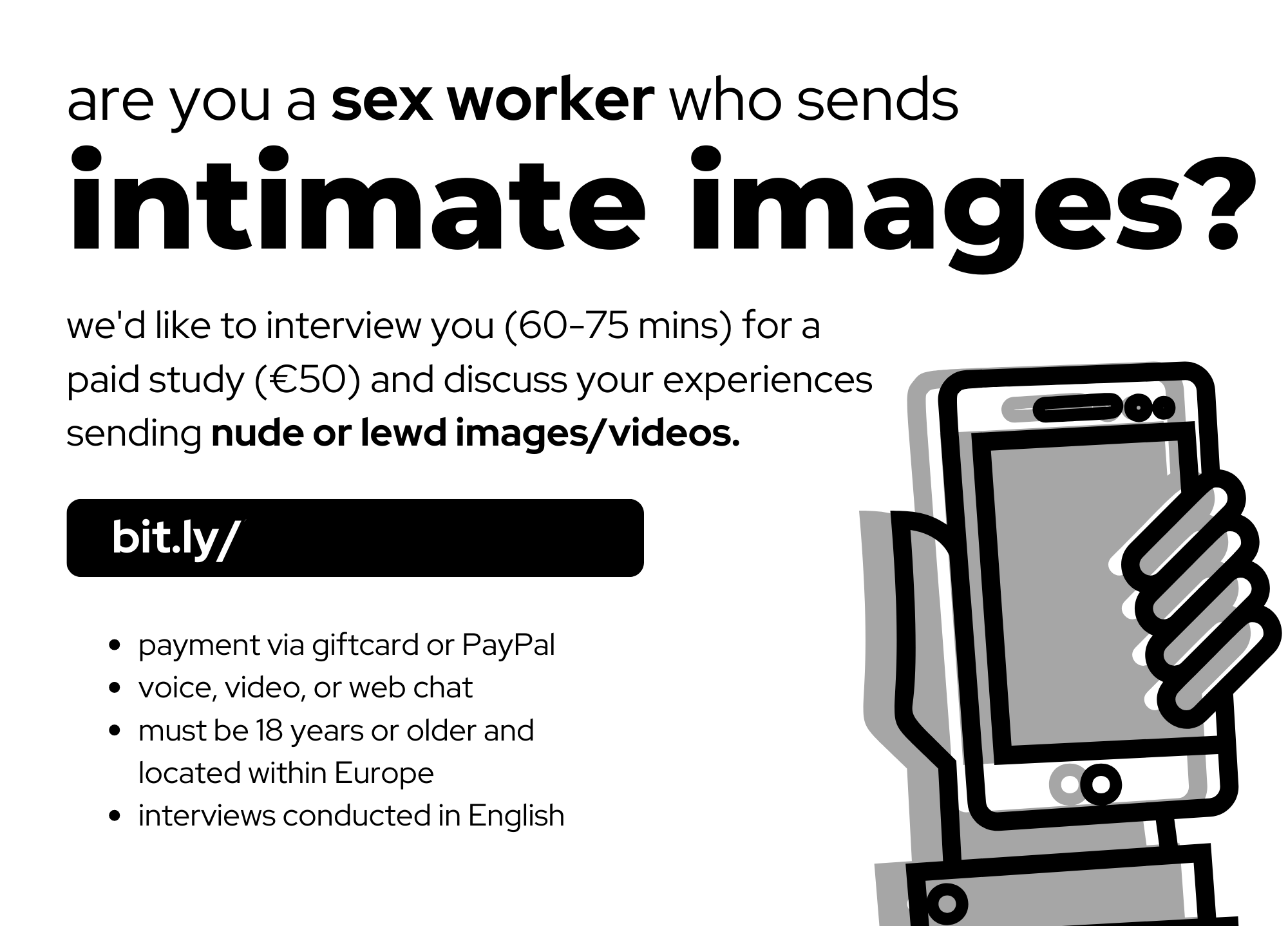}
\caption{Recruitment graphic for sex workers}
\end{figure}

\subsection{Interview Protocols}
\label{appendix:questions}
Our interview protocols are available at the following link \edit{along with our codebook}: 
$$
\texttt{bit.ly/Tech-IBSA-Research-Protocols}
$$

The audio and video interviews lasted an average of 47 minutes. During the interviews, we explicitly asked participants about the usefulness of a selected set of features based on prior work and existing media advice guides. These features included: end-to-end encryption, expiring messages, screenshot notifications, digital fingerprinting, watermarking, photo editing tools, and metadata removal. We wrote a high-level explanation of each feature that interviewers read to participants who had not previously heard of them or were only vaguely familiar with them. A document containing these explanations are included in the same repository.

\end{document}